\documentclass[11pt]{llncs}
\usepackage[letterpaper,hmargin=1.00in,vmargin=1.00in]{geometry}
\usepackage{amssymb}
\usepackage{color}
\newcommand{\ignore}[1]{} \newcommand{\full}[1]{} 
\newcommand{\rand}{\stackrel{R}{\leftarrow}}
\newcommand{\la}{\left\langle}
\newcommand{\ra}{\right\rangle}
\newcommand{\mc}{\mathcal}
\newcommand{\ms}{\texttt}
\newcommand{\mb}{\mathbb}

\newcommand{\lea}{\leftarrow}
\newcommand{\poly}{\mathbb{P}}
\newcommand{\alf}{\mathbb{ALF}}
\newcommand{\spec}{\mathbb{SPEC}}
\newcommand{\lfm}{\mathbb{LF}}
\newcommand{\tm}{\mathbb{TM}}
\newcommand{\ptmf}{\mathbb{PTMF}}
\newcommand{\pptmf}{\mathbb{PPTMF}}
\newcommand{\tmf}{\mathbb{TMF}}
\newcommand{\alforlf}{(\mathbb{A})\mathbb{LF}}
\newcommand{\ptm}{\mathbb{PTM}}
\newcommand{\ppt}{\mathbb{PPT}}

\usepackage{hyperref}

\author{Amitabh Saxena$^a$ and Brecht Wyseur$^{b}$\\$~$\\
        {$^a$}DISI, University of Trento, Italy\\
        amitabh@disi.unitn.it\\$~$\\
        {$^{b}$}ESAT/SCD -- COSIC/IBBT, K.U. Leuven, Belgium\\
        brecht.wyseur@esat.kuleuven.be
        }
\title{On White-Box Cryptography and Obfuscation\thanks{This work was partly supported by funds from the European Commission through the IST Program under Contract IST-021186-2 for the RE-TRUST project and in part by the IAP Program P6/26 BCRYPT of the Belgian State (Belgian Science Policy).}}
\institute{~}%RE-TRUST Project\\www.re-trust.org}
\def\keywordsname{Keywords.}
\newenvironment{keywords}{%
      \list{}{\advance\topsep by-0.50cm\relax\small
      \leftmargin=1cm
      \labelwidth=1cm%\z@
      \listparindent=1cm%\z@
      \itemindent\listparindent
      \rightmargin\leftmargin}\item[\hskip\labelsep
                                    \bfseries\keywordsname]}
    {\endlist}
\let\oldmarginpar\marginpar
\renewcommand\marginpar[1]{\oldmarginpar{\footnotesize #1}}

\renewcommand\marginpar[1]{}
\newcommand\amit[1]{\color{red} #1}
\newcommand\brecht[1]{\color{magenta} #1}
\newcommand\greg[1]{\color{brown} #1}
\newcommand\todo[1]{\color{green} #1}
\begin{document}
\maketitle
\marginpar{\underline{Coloring legend}
\begin{itemize}
 \item {\color{red}Amitabh}
 \item {\color{magenta}Brecht}
 \item {\color{brown}Gregory}
 \item \todo{todo}
\end{itemize}
}

\marginpar{{\brecht Margins etc. adjusted to CRYPTO'08 requirements}
{\todo
\begin{enumerate}
%	\item Give existing defns of obfuscators (Hohenberger et al.'s~\cite{Hohenberger07re-encryption}, Barak's~\cite{Barak2001impossibility}, others?)
	\item Prove that UWBP can be satisfied for non-learnable but approximately learnable functions (e.g., point functions) based on some computational assumption (such as discrete log problem). Specifically, finish proof of Theorem~\ref{uwbpforalf}

	\item Give criteria to decide of WBP is satisfied for $(Q, spec)$ by a given obfuscator. (point 3 in ``Our contribution'' Section). This will come in Section~\ref{wbpfromobf}.
	
	\item Insert citations  [to include: Goldwasser, Barak, Hohenberger, Dennis, Wee, Canetti (2008), BF-IBE, others?]
	\item If necessary, write ``related work'' section, describe existing works.
	\item Elaborate/clarify ``specification'', ``experiment'', and any other newly introduced concepts.
%	\item Av. case secure obfuscation: is it comparable to our defns? Its based on randomized functions, while we consider deterministic functions.
%	\item finish proofs of Lemmas 1, 2, 3, 5, 7 and 8. (1, 2: Trivial. 3, 5: Medium. 7, 8: Hard). Note: 7 is already proved in~\cite{wee05obfuscating}. Perhaps we can use it if its based on IND definitions.
\end{enumerate}
}}

\begin{abstract}

We study the relationship between obfuscation and white-box cryptography. We capture the requirements of any white-box primitive using a \emph{White-Box Property (WBP)} and give some negative/positive results. Loosely speaking, the WBP is defined for some scheme and a security notion (we call the pair a \emph{specification}), and implies that w.r.t. the specification, an obfuscation does not leak any ``useful'' information, even though it may leak some ``useless'' non-black-box information.

Our main result is a negative one - for most interesting programs, an obfuscation (under \emph{any} definition) cannot satisfy the WBP for every specification in which the program may be present. To do this, we define a \emph{Universal White-Box Property (UWBP)}, which if satisfied, would imply that under \emph{whatever} specification we conceive, the WBP is satisfied. We then show that for every non-approximately-learnable family, there exist certain (contrived) specifications for which the WBP (and thus, the UWBP) fails. 

On the positive side, we show that there exists an obfuscator for a non-approximately-learnable family that achieves the WBP for a certain specification. Furthermore, there exists an obfuscator for a non-learnable (but approximately-learnable) family that achieves the UWBP.

Our results can also be viewed as formalizing the distinction between  ``useful'' and ``useless'' non-black-box information.

\end{abstract}


%irrespective of whether it leaks . We give a separation between useful and non-black-box information. That is, even if the obfuscation leaks some non-black-box information,
%Our main result is a negative one - under \emph{whatever} definition of obfuscation we use, for most programs there will be cerhe idea of useful information
%lies that

\section{Introduction}

%\subsection{Obfuscation}
%{\scshape{Obfuscation.}}
Informally, an obfuscator $O$ is a probabilistic compiler that transforms a program $P$ into $O(P)$, an executable implementation of $P$ which hides certain functional characteristics of $P$. Starting from the seminal work of Barak {\em et al.}~\cite{Barak2001impossibility}, several definitions for obfuscators have been proposed~\cite{HofheinzMS07,Hohenberger07re-encryption,wee05obfuscating}, each one based on some sort of {\em virtual black-box property (VBBP)}. Loosely speaking, the VBBP requires that {\em whatever we could do using the obfuscated program, we could also have done using black-box access to the original program}. The notion of ``whatever'' can be captured using several formalisms. The following are the common ones (in decreasing order of generality):

\begin{enumerate}
%	\item Perfect indistinguishability
	%\item Computational indistinguishability (Wee~\cite{wee05obfuscating}, Barak~\cite{Barak2001impossibility}, Hohenberger~\cite{Hohenberger07re-encryption}, Hofheinz~\cite{HofheinzMS07})
	\item Computing something that is indistinguishable from the obfuscation~\cite{Barak2001impossibility,HofheinzMS07,Hohenberger07re-encryption,wee05obfuscating}.
	\item Computing some function~\cite{Barak2001impossibility}.
	%\item Computing some fixed function [??]
	\item Computing some predicate~\cite{Barak2001impossibility}.
	%\item Computing some fixed predicate [??]
\end{enumerate}
\marginpar{Insert related work here}

\subsection{White-box Cryptography}

\emph{White-box cryptography (WBC)}, which requires that \emph{some given scheme must remain secure even if the adversary is given ``white-box access'' to a functionality instead of just black-box access}, is an active field of research. Informally, white-box access implies that the adversary is given an executable implementation of the algorithm that was used inside the black-box~\cite{chow02wbaes,chow02wbdes}. Existing notions of WBC only deal with the encryption algorithm of symmetric block-ciphers. In this work, we generalize this intuition to any cryptographic primitive. For instance, we can use WBC to convert a MAC into a signature scheme by white-boxing the verification algorithm.

{\scshape{White-Box Security.}} The (black-box) security of any primitive is captured using a \emph{security notion} (e.g., IND-CPA) where the adversary is given black-box access to some functionality (e.g., encryption), and a white-box implementation can be required to satisfy that security notion when the adversary is given access to a \emph{white-boxed} version of the functionality.

\subsection{Motivation}

One way to realize WBC is to obfuscate (using an obfuscator) the executable code of the algorithm and hope that the adversary cannot use it in a non-black-box manner. What we would like is, given an obfuscator satisfying some definition, a white-box implementation can be proved secure under some security notion. Furthermore, if a scheme is required to satisfy several security notions simultaneously (Authenticated Encryption (AE)~\cite{bellare-auth} and the Obfuscated Virtual Machine (OVM) of~\cite{cryptoeprint:2008:087} are two such examples, where both confidentiality and integrity needs to be satisfied),
%WBRPC must satisfy integrity and confidentiality) \marginpar{discard WBRPC; other example needed. E.g., we could use Authenticated Encryption or sign-encryption, which involves both encryption and authentication.},
we would like the obfuscation to ensure that all the security notions are satisfied in the white-box variant if they are satisfied in the black-box variant. %However, it is still not fully clear if the any of the existing definitions of obfuscators can be used to achieve these goals. Hence, a natural question is:
However, it is still not fully clear if any of the existing definitions of obfuscators can be used to achieve these goals. Hence, a natural question is:

\begin{enumerate}
	\item []\emph{Given an obfuscator satisfying the virtual black-box property for a program $P$ (in some sense), and some scheme that is secure when the adversary is given black-box access to $P$, can it be proved (without additional assumptions) that the scheme remains secure when the adversary is also given access to the obfuscated program $O(P)$?}
\end{enumerate}

\subsection{Our Contribution}
%\marginpar{\greg{The ``main result'' mentioned below can be a starting observation for your research, but is certainly not a result by itself}}
\begin{enumerate}
	\item In this paper we answer the above question in the negative - we show that under \emph{whatever} definition of obfuscation we use, the answer to the above question is, in general, no.
	To do this, we first define the objective(s) of a white-box primitive, which we formalize using a \emph{white-box property (WBP)}. %\marginpar{\color{brown}``This can be a starting observation for your research, but is certainly not a result by itself''}
	Our main observation is that when considering obfuscation of most programs $P$, we must also take into account the scheme plus the security notion (i.e., the specification) in which $P$ is used. Furthermore, we show that for most programs $P$, there cannot exist an obfuscator that satisfies the WBP for all specifications in which $P$ might be present. To do this, we define a \emph{universal white-box property (UWBP)} which, if satisfied, would imply that in whatever specification $P$ might be present, the obfuscated program $O(P)$ will not leak any ``useful'' information. We then show that for every non-approximately-learnable program $P$, there exists some specification in which the obfuscation leaks useful information, thereby failing the UWBP. %Our counter-examples are based on those the work of~\cite{Barak2001impossibility}.
\item On the positive side, we have the following two results.
\begin{enumerate}
	\item We show that under reasonable computational assumptions, there exists an obfuscator that satisfies the WBP w.r.t. some meaningful specification for a non-approximately-learnable program $P$.
	\item We show that there exist obfuscators that satisfy UWBP for a program that is non-learnable but approximately learnable. %This result is based on the obfuscation of point functions.
\end{enumerate}
%\item Finally, we give a method to determine if the white-box property for some given specification can be reduced to simulation-based definitions of obfuscation (such as Hohenberger {\em et al.}'s~\cite{Hohenberger07re-encryption}).
\end{enumerate}

\section{Related Work}

Practical white-box implementations of DES and AES encryption algorithms were proposed in~\cite{chow02wbaes,chow02wbdes}. However, no definitions of obfuscation were given, neither were there any proofs of security. With their subsequent cryptanalysis~\cite{billet04wbaes,goubin07wbdes,wyseur07wbdes}, it remains an open question whether or not such white-box implementations exist.

The notion of code-obfuscation was first given by Hada in~\cite{Hada00ZKobfus}, which introduced the concept of virtual black-box property (VBBP) using computational indistinguishability. In~\cite{Barak2001impossibility}, Barak \emph{et al.} defined obfuscation using the weaker predicate-based VBBP and showed that there exist unobfuscatable function families under their definition. Goldwasser and Kalai~\cite{GK05} extend the impossibility results of~\cite{Barak2001impossibility} w.r.t. auxiliary inputs.

On the positive side, there have been several results too. For instance, Lynn \emph{et al.} show in~\cite{lynn04positive} how to obfuscate point functions in the random oracle model. Wee in~\cite{wee05obfuscating} showed how to obfuscate point functions without random oracles. Hohenberger \emph{et al.}~\cite{Hohenberger07re-encryption} used a stronger notion of obfuscation (average-case secure obfuscation) and showed how it can be used to prove the security of re-encryption functionality in a weak security model (i.e., IND-CPA). They also presented a re-encryption scheme under bilinear complexity assumptions. Hofheinz \emph{et al.}~\cite{HofheinzMS07} discuss a related notion of obfuscation and show that IND-CPA encryption and point functions can be securely obfuscated in their definition. Goldwasser and Rothblum~\cite{DBLP:conf/tcc/GoldwasserR07} define the notion of ``best-possible obfuscation'' in order to give a qualitative measure of information leakage by an obfuscation (however, they do not differentiate between ``useful'' and ``useless'' information). Recently, Canetti and Dakdouk~\cite{canetti08multibit} give an obfuscator for point functions with multi-bit output for use in primitives called ``digital lockers''. Finally, Herzberg \emph{et al.}~\cite{cryptoeprint:2008:087} introduce the concept of \emph{White-Box Remote Program Execution (WBRPE)} in order to give a meaningful notion of ``software hardening'' for all programs and avoid the negative results of~\cite{Barak2001impossibility}.

However, till date, there has not been much work done on the relationship between arbitrary white-box primitives and obfuscation. This paper is intended to fill this gap.

\section{Preliminaries}

\marginpar{This should eventually be truncated, or perhaps partially moved to an Appendix, and should also cite to papers that cover these definitions}

Denote by $\poly$ the set of all positive polynomials and by $\tm$ the set of all Turing Machines (TMs). All TMs considered in this paper are deterministic (a probabilistic TM is simply a deterministic TM with randomness on the input tape). A mapping $f:x\ni\mb{N}\mapsto f(x)\in \mb{R}$ is negligible in $x$ (written $f(x)\leq negl(x)$) if $\forall p\in\poly,\exists x'\in \mb{N},\forall x>x':f(x)< 1/p(x)$.

For simplicity, we define the input-space of arbitrary TMs to be $\{0,1\}^*$, the set of all strings. If, however, the input-space of a TM is well defined and efficiently samplable (for instance, the strings should be of a particular encoding), then we implicitly imply that the inputs are chosen from the input-space sampled using a string from $\{0,1\}^*$. All our definitions and results apply in this extended setting without any loss of generality.

\begin{definition}\label{big-defn1} In the following, unless otherwise stated, a TM is assumed to have only one input tape.
\begin{enumerate}
	\item \emph{\bfseries(Equality of TMs.)} $X, Y\in \tm$ are equal (written $X = Y$) if $\forall a:X(a)=Y(a)$
%	\item \emph{\bfseries(Approximate Equality of TMs.)} $X, Y\in\tm$ are approximately equal (written $X\approx Y$) if $\forall \ell:\Pr[a\rand \{0,1\}^\ell:X(a)= Y(a)]\geq 2/3$. (The constant $2/3$ is arbitrary. However a constant closer to 1 makes our impossibility results of \S\ref{negative-results} stronger.)

%Note that the above definition of approx. equality is very weak (in fact too weak to be of any practical use). However, our negative results of \S\ref{negative-results} are the strongest in this weakest formulation.
	%For any TMs $A, B, C$, if $A\approx B$ and $B=C$, then $A\approx C$.

%	\begin{remark} %It might seem that the parameter $1/\ell$ on the RHS above is a bit ad-hoc, which is true.
%	\emph{What we would ideally like to capture is that ``for randomly chosen inputs $a$, the probability that $X(a)\neq Y(a)$ is very small.'' However, it turns out that there is no meaningfully way define approx. equality for arbitrary TMs. On the other hand, the parameter $1/\ell$ turns out to be suitable for the purposes of our counter-example of \S\ref{negative-results}. }
%	
%	\end{remark}
%	\begin{remark} \emph{Both $=$ and $\approx$ are equivalence relations: If $X=Y$ and $Y=Z$ then $X=Z$. Similarly, if $X\approx Y$ and $Y\approx Z$ then $X\approx Z$.}
%	\end{remark}
	%\footnote{Note that for TMs, both $=$ and $\approx$ are equivalence relations.}
	\item \emph{\bfseries(Polynomial TM.)} $X\in\tm$ is a Polynomial TM (PTM) if there exists $p\in\poly$ s.t. $\forall a:X(a)$ halts in at most $p(|a|)$ steps. Denote the set of all PTMs by $\ptm$.
		\item \emph{\bfseries(PPT Algorithms.)} A PPT algorithm (such as an adversary or an obfuscator) is a PTM with an unknown source of randomness input via an additional random tape. We denote the set of PPT algorithms by $\ppt$. The running time of a PPT algorithm must be polynomial in the length of the known inputs.

	\item \emph{\bfseries(TM Family.)} A TM Family (TMF) is a TM having two input tapes: a \emph{key tape} and a \emph{standard input tape}. We denote by $\tmf$ the set of all TMFs. Let $Q\in \tmf$. Then:%\footnote{Here $Q$ can be considered as describing a function ensemble (such as the RSA encryption algorithm), and the key tape can be considered as containing the key.}	
\begin{enumerate}
	\item 	The symbol $Q^q$ indicates that the key tape of $Q$ contains string $q$. %\footnote{For notational convenience, we will write the key in superscript instead of the usual subscript notation.}
	\item 	We denote by $\mc{K}_Q$ the key-space (valid strings for the key tape) of $Q$. %We denote by $\tmf$ the set of all TMFs.
	\item   Let $q\in\mc{K}_Q$. In our model, the input-space (valid strings for the standard input tape) of $Q^q$ is fully defined by the parameter $|q|$. We denote this space by $\mc{I}_{Q, |q|}$. Furthermore the following must hold: 
	\[\exists p\in \poly, \forall q \in \mc{K}_Q,\forall x\in \mc{I}_{Q, |q|}:|x|=p(|q|).\] %(shorter inputs are padded).% This can be easily done for shorter inputs
%\begin{enumerate}
%	\item $\exists p\in \poly,\forall q\in \mc{K}_Q:|\mc{I}_{Q,|q|}|\leq 2^{p(|q|)}$.% for some $p\in \poly$. %the size of the key-space and the input-space must be polynomially related. For convenience, however, we will assume that their sizes are the same.
%	\item $\exists p\in \poly, \forall q \in \mc{K}_Q,\forall x\in \mc{I}_{Q, |q|}:|x|=p(|q|)$. %(shorter inputs are padded).% This can be easily done for shorter inputs using appropriate padding.
%\end{enumerate}
\end{enumerate}
	\item \emph{\bfseries(Polynomial TM Family.)} $Q\in\tmf$ is a \emph{Polynomial} TMF (PTMF) if there exists $p\in\poly$ such that $\forall q\in \mc{K}_{Q},\forall a\in\mc{I}_{Q,|q|}: Q^q(a)$ halts in at most $p(|q|)$ steps. %$p(|q|+|a|)$ steps
	We denote the set of all PTMFs by $\ptmf$.
%\end{description}	

%\begin{description}
	\item \label{lf} \emph{\bfseries(Learnable Family.)} $Q\in\tmf$ is learnable if $\exists (L, p)\in\ppt\times \poly$ s.t. \[\forall k:\Pr[q\rand\{0,1\}^k\cap \mc{K}_Q;X\lea L^{Q^q}(1^{|q|},Q):X= Q^q]\geq 1/p(k)\] (the probability taken over the coin tosses of $L$) and:
\begin{enumerate}
	\item $\forall a:$ if $Q^q(a)$ halts after $t$ steps then $X(a)$ halts after at most $p(t)$ steps.\footnote{This condition is to prevent an exponential time learner from becoming polynomial time by hard-wiring the learning algorithm and queries/responses inside $X$.}
	\item $|X|\leq p(|q|)$.
\end{enumerate}
	$L$ is called the learner for $Q$. We denote the set of all learnable families by $\lfm$.
	\item \label{alf}\emph{\bfseries(Approx. Learnable Family.)} $Q\in\tmf$ is approx.  learnable if $\exists (L,p)\in\ppt\times\poly$ s.t. \[\forall k:\Pr[q\rand\{0,1\}^k\cap \mc{K}_Q;a\rand\mc{I}_{Q, k};X\lea L^{Q^q}(1^{|q|},Q):X(a)= Q^q(a)] \geq 1/p(k)\] (the probability taken over the coin tosses of $L$), and: 
\begin{enumerate}
	\item $\forall a:$ if $Q^q(a)$ halts after $t$ steps then $X(a)$ halts after at most $p(t)$ steps.
	\item $|X|\leq p(|q|)$.
\end{enumerate}
 %\footnote{A PTMF may be approximately learnable but not learnable.}
We denote the set of all approx. learnable families by $\alf$.
\end{enumerate}
\end{definition}
\begin{lemma} \label{equivalence-between-non-learn-families} If $Q_1\in \ptmf\backslash \alforlf$, then the following holds:%\footnote{%Lemma~\ref{equivalence-between-non-learn-families} says: Let $Q_1 \in \ptmf\backslash \alforlf$ and let there exist a polynomial $p$ and $Q_2\in \ptmf$ such that for every $q_1$ there exists $q_2$ with $|q_2|\leq p(|q_1|)$ and $Q_1^{q_1}= Q_2^{q_2}$, then $Q_2\notin \alforlf$.
%$(\mathbb{A})$ indicates that $\mathbb{A}$ is optional in the above statement. %The lemma requires that such a $Q_2$ to only exist, and not necessarily be known.
%}
\[\Big[\exists (Q_2,p)\in \ptmf\times \poly,\forall q_1\in \mc{K}_{Q_1},\exists q_2\in\mc{K}_{Q_2}:Q_1^{q_1}=Q_2^{q_2}\wedge |q_2|\leq p(|q_1|)\Big] \rightarrow Q_2 \notin \alforlf.\]
The symbol $(\mathbb{A})$ indicates that $\mathbb{A}$ is optional in the above statement.
\end{lemma}
\begin{proof} Assume for contradiction that for any given $Q_1\in \ptmf$ that is not learnable, there exists some $(Q_2, p)\in \ptmf\times poly$ such that the LHS of the above implication is satisfied but RHS is not. Let $L_1$ and $L_2$ be the learners for $Q_1$ and $Q_2$ respectively. %(Although $L_1$ may not know the description of $Q_2$ (and so does not know the description of $L_2$, the learner for $Q_2$), we may assume that $|Q_2|$ (the length of the description of $Q_2$) is upper-bounded by $c$ for some constant $c$. This implies that there are at most $2^{c}$ possible candidates for $L_2$. In this case for any given key $q_1$ and oracle access to $Q_1^{q_1}$, $L_1$ first does a brute-force search on these $2^{c}$ different instances of $Q_2$ (i.e., of $L_2$) and outputs whatever the correct $L_2$ outputs. The running time of $L_1\leq2^{c}\cdot c'\cdot(\max($running time of $L_2))$, where $c'$ is a constant indicating the time to test each instance during the brute-force search. Hence we arrive at a contradiction.
$L_1$ runs $L_2$ using its own oracle to answer $L_2$'s queries. If $Q_2$ is learnable, then $L_2$ will output $X_2\approx Q_2^{q_2}$ in a polynomial (of $|q_2|$) number of steps, which is a polynomial function of $|q_1|$ by assumption, a contradiction.
\qed\end{proof}

\section{Obfuscators}
\label{obfuscators}
%\marginpar{\amit{We give a new definition of obfuscation based on the semantic security property of encryption schemes. For comparison, we also present various existing definitions.
%In the vein of conventional cryptography, we define the functionality of the obfuscator using a {\em correctness} property and the security using a {\em soundness} property. We give a single formulation of correctness (which is more or less the same in all definitions) and various formulations of soundness, each corresponding to a different definition of obfuscation. }}
%\marginpar{\amit{Note: the new definition we intended to give was UFVBBP. However, if we remove it or decide that thats not the main focus of this paper, we can rewrite the above paragraph and say that this is a summary of existing definitions (and not a new definition)}}

In this work, we only consider obfuscation of PTMFs with a uniformly selected key, and not of a single PTM. %\footnote{We argue that a meaningful notion of obfuscation cannot exist for a single PTM, since it is not clear how this PTM is selected.}
As is common in cryptography, we define the functionality of the obfuscator using a {\em correctness} property and the security using a {\em soundness} property. In contrast to existing works, however, we define an obfuscator using only the correctness property. This is to consider different notions of ``white-box'' security (which might be unrelated to soundness) and still be able to use the word ``obfuscator'' in a formal sense.

\subsection{Obfuscator (Correctness)}
\label{correctness}
\begin{definition} \label{correctness-defn} A randomized algorithm $O:\ptmf\times\{0,1\}^*\mapsto \tm$ satisfies \emph{correctness} for $Q\in \ptmf$ if  the following two properties are satisfied:%\footnote{For this paper, we only consider obfuscation of PTM families. We argue that a meaningful notion of obfuscation cannot be obtained for a single TM.}
\begin{enumerate}
	\item \label{apf}\textbf{Approx. functionality:}
		\[\forall q\in \mc{K}_Q,\forall a\in \mc{I}_{Q, |q|}:\Pr[O(Q, q)(a)\neq Q^q(a)]\leq negl{(|q|)},\]
%	{\color{magenta}Shouldn't it be $\leq negl(|a|)$?}
%	{\color{red} No. we want it to be poly in key length. This automatically implies that exponential length $a$ will be truncated and only poly bits will be used. The key idea of our paper is the notion of PTMFs. We could add $negl(|q|+|a|)$, but I feel that defining all parameters in terms of key length only has many advantages. }
	 the probability taken over the coin tosses of $O$.\footnote{For now, we consider the functionality of $Q$ only in a deterministic sense. That is, we do not consider the notion of obfuscation of ``probabilistic functions'' (used, for example, in~\cite{HofheinzMS07,Hohenberger07re-encryption}). However, our negative results (presented in \S\ref{negative-results}) also apply to probabilistic functions using an appropriately defined notion of \emph{probabilistic PTMFs (PPTMFs)} (and a corresponding notion of approx. functionality for PPTMFs). This aspect will be further discussed in~\S\ref{pptmfs}.}%, where the definition of approximate functionality is extended to that of~\cite{HofheinzMS07,Hohenberger07re-encryption}. %However, we leave out the latter analysis for simplicity.

	 \item \textbf{Polynomial slowdown and expansion:} There exists $p\in\poly$ s.t.
	 \[\forall q\in  \mc{K}_Q:|O(Q, q)|\leq p(|q|),\]
	and $\forall a$, if $Q^q(a)$ halts in $t$ steps then $O(Q, q)(a)$ halts in at most $p(t)$ steps.
	\end{enumerate}
We say the $O$ is efficient if $O\in\ppt$.

If $O$ satisfies correctness for $Q$, we say that $O$ is an \emph{obfuscator} for $Q$.
%\end{definition}
\marginpar{\amit For simplicity, we only consider TMs in this paper. However, all our results also carry over to circuits (to check this last statement!)}
\end{definition}

%\begin{lemma}\label{approx-func-implies-approx-equiv}Let $Q\in \ptmf$ and let $O$ be an obfuscator for $Q$. Then for sufficiently large $k$, and every $q\in \mc{K}_Q\cap \{0,1\}^k: Q^q\approx O(Q,q)$.\end{lemma}
%\begin{proof} Under Definition~\ref{correctness-defn}, for sufficiently large $k$ the following is guaranteed to hold:\[\forall a:\Pr[O(Q, q)(a)=Q^q(a)]\geq 1/3.\] %The approx. functionality requirement is strictly stronger than the approx. equality requirement because the former holds for every $a$, while the latter is required to hold only for uniform $a$. %, and (b) negligible in $(|q|+|a|)$ implies negligible in $|a|$.
%Hence, the lemma follows.
%\qed\end{proof}%
%\marginpar{\amit{\begin{proof} (Lemma~\ref{approx-func-implies-approx-equiv}) The approximate functionality requirement is strictly stronger than the approximate equality requirement. (because the former holds for every $q, a$, while the latter is required to hold only for uniformly chosen $q, a$). Hence, the result follows.\end{proof}
%}}

\subsection{Obfuscator (Soundness)}
\label{soundness}
Over recent years, several definitions of soundness have been proposed, all based on some sort of {\em Virtual Black-Box Property (VBBP)}~\cite{Barak2001impossibility,HofheinzMS07,Hohenberger07re-encryption,lynn04positive,wee05obfuscating}. Let $Q\in \ptmf$ and let $q\in \{0,1\}^*$. Loosely speaking, the VBBP requires that whatever information about $q$ a PPT adversary computes given the obfuscation $O(Q, q)$, a PPT simulator could also have computed using only black-box access to $Q^q$. All existing notions of VBBP can be classified into one of two broad categories. At one extreme (the weakest) are the predicate-based definitions, where the adversary and the simulator are required to compute some predicate of $q$. At the other extreme (the strongest) are definitions based on computational indistinguishability, where the simulator is required to output something that is indistinguishable from $O(Q, q)$. We define these two notions below. Our definitions are based on that of~\cite{GK05}, where an auxiliary input is also considered.

%\subsubsection{Virtual Black-Box Property (VBBP)}
\begin{definition} \label{soundness-defn} An obfuscator $O$ for $Q\in \ptmf$ satisfies \emph{soundness} for $Q$ if at least one of the properties given below is satisfied.

\begin{enumerate}
	
%<<<<<<< .mine
	\item \textbf{Predicate Virtual black-box property (PVBBP):} Let $\pi$ be any efficiently verifiable %(but not necessarily efficiently computable) 
	predicate on $\mc{K}_Q$. $O$ satisfies PVBBP for $Q$ if 
%=======
%	\item \textbf{Predicate Virtual black-box property (PVBBP):} Let $\pi$ be any predicate on $\mc{K}_Q$. $O$ satisfies PVBBP for $Q$ if
%>>>>>>> .r115
	\[\forall (A,p)\in\ppt\times\poly,\exists (S, k')\in\ppt\times\mathbb{N},\forall k>k':Adv^{pvbbp}_{A, S, O, Q}(k)\leq negl(k),\]
	
	 where
%	\[Adv^{pvbbp}_{A, S, O, Q}(k)=\max_{z\in \{0,1\}^{p(k)}}\left|\begin{array}{c}\Pr[{q\rand \{0,1\}^{k}}\cap \mc{K}_Q:A^{Q^q}(1^k, O(Q, q), z)=1]\\-\Pr[{q\rand \{0,1\}^{k}\cap \mc{K}_Q}:S^{Q^{q}}(1^k,z)=1 ]\end{array}\right|,\]
	\[Adv^{pvbbp}_{A, S, O, Q}(k)=\max_{\pi}\max_{z\in \{0,1\}^{p(k)}}\left|\begin{array}{c}\Pr[{q\rand \{0,1\}^{k}}\cap \mc{K}_Q:A^{Q^q}(1^k, O(Q, q),z)=\pi(q)]\\-\Pr[{q\rand \{0,1\}^{k}\cap \mc{K}_Q}:S^{Q^{q}}(1^k,z)=\pi(q) ]\end{array}\right|,\]
	the probability taken over the coin tosses of $O, A, S$.\footnote{The definition of PVBBP given here is slightly weaker than the one used in~\cite{Barak2001impossibility} because they require this property to hold for every $q$, while we require it to hold only for uniformly selected $q$.}

	\item \textbf{Computational Indistinguishability (IND):} $O$ satisfies IND for $Q$ if
	\[\forall (A,p)\in\ppt\times\poly,\exists (S,k')\in\ppt\times\mathbb{N},\forall k>k': Adv^{ind}_{A, S, O, Q}(k)\leq negl(k),\] where	
	%\[Adv^{ind}_{A, S, O, Q}(k)=\max_{z\in \{0,1\}^{p(k)}}\left|\begin{array}{c}\Pr[{q\rand \{0,1\}^{k}\cap \mc{K}_Q}:A^{Q^q}(1^k, O(Q, q),z)=1]\\-\Pr[{q\rand \{0,1\}^{k}\cap \mc{K}_Q}:A^{Q^q}(1^k,S^{Q^q}(1^k),z)=1 ]\end{array}\right|,\]
	\[Adv^{ind}_{A, S, O, Q}(k)=\max_{z\in \{0,1\}^{p(k)}}\left|\begin{array}{c}\Pr[{q\rand \{0,1\}^{k}\cap \mc{K}_Q}:A^{Q^q}(1^k, O(Q, q),z)=1]\\-\Pr[{q\rand \{0,1\}^{k}\cap \mc{K}_Q}:A^{Q^q}(1^k,S^{Q^q}(1^k,z),z)=1 ]\end{array}\right|,\]
	the probability taken over the coin tosses of $O, A, S$.
\end{enumerate}
Depending on the property satisfied, we call it IND-soundness or PVBBP-soundness (note that the former implies the latter).
\end{definition}%
%\begin{definition}
%
%A randomized algorithm $O:\ptmf\times\{0,1\}^*\mapsto \tm$ is an obfuscator for $Q\in \ptmf$ if it satisfies correctness and soundness for $Q$. %
%%Depending on the property satisfied, we call it IND-soundness or PVBBP-soundness (note that the former implies the latter).
%%
%%If $O$ satisfies correctness and soundness for $Q$, we say that $O$ is an obfuscator for $Q$.
%\end{definition}

It has been noted (but never proved) in several papers (e.g.,~\cite{Barak2001impossibility,Hohenberger07re-encryption}) that the PVBBP is too weak for practical purposes. Furthermore, it has been noted that the IND-soundness is too strong to be satisfied in practice~\cite{Hohenberger07re-encryption,wee05obfuscating}. In fact, it is easy to prove:

\begin{proposition} \label{ind-to-alf} If there exists an obfuscator satisfying IND-soundness for some $Q\in \ptmf$ then $Q\in \alf$.\footnote{This result does not hold if the definition of approx. functionality in correctness is extended to probabilistic functions. See \S\ref{pptmfs} for details.}
\end{proposition}

Nevertheless, it is conceivable that a definition of soundness can be formulated falling somewhere between the two extremes, which is neither too weak nor too strong, and can be used for proving security of arbitrary white-box primitives. We show this is not the case. Specifically, we show that, under \emph{whatever} definition of soundness we use, for every family $Q\notin \alf$, there exist (contrived) specifications for which white-box security fails but the corresponding black-box construction is secure.\footnote{In related work, the authors of~\cite{HofheinzMS07} show that a slightly different notion of the IND property - one based on probabilistic functions - is insufficient for proving the white-box IND-CCA2-security of encryption schemes, even if white-box IND-CPA is satisfied. Our results are more general because they apply to every $Q\notin \alf$.}% Although we only consider obfuscation of deterministic functions, our results can be carried over to probabilistic functions (e.g., of~\cite{HofheinzMS07,Hohenberger07re-encryption}). %However, we leave out the latter analysis for simplicity.

\section{White-box Cryptography (WBC)}
\label{white-box-cryptography}

In this section, we formalize the notion of WBC by defining a \emph{white-box property (WBP)}. A key concept of our model is the notion of a (cryptographic) \emph{specification}. Informally, a specification is a self-contained description (in some formal language) of a cryptographic scheme (such as RSA-OAEP) along with a corresponding security notion (such as IND-CPA). We follow the basic principles of various ``game-based'' approaches~\cite{bellare98puk,DBLP:conf/eurocrypt/BellareR06,DBLP:conf/stoc/GoldwasserM82,DBLP:journals/jcss/GoldwasserM84} where a security notion  is captured using an interactive game between an adversary and a challenger. In our model, the role of the challenger is played by an \emph{experiment} and the corresponding game is called a \emph{simulation}. We denote by $\spec$ the set of all specifications. %%attempt to formalize this intuition. %A formal treatment of this notion follows.

\subsection{Black-Box Simulation}
\label{black-box-sim}
Let $spec\in\spec$ denote the specification of some scheme (e.g., ``IND-CPA security notion for symmetric encryption scheme X''). Every such $spec$ defines a {\em Black-box simulation} (or simply simulation) between an \emph{experiment} and an \emph{adversary}.

\textbf{Experiment.} The experiment for $spec$, written $\texttt{Expt}^{spec}$ is a TM having six tapes: (1) a read-only {\em experiment-input} tape, (2) a writable {\em adversary-input} tape, (3) a read-only {\em query-input} tape, (4) a writable {\em query-response} tape, (5) a read-only {\em adversary-output} tape, and (6) a writable {\em experiment-output} tape
%\begin{enumerate}
%	\item A read-only {\em experiment-input} tape
%	\item A writable {\em adversary-input} tape
%	\item A read-only {\em query-input} tape
%	\item A writable {\em query-response} tape
%	\item A read-only {\em adversary-output} tape
%	\item A writable {\em experiment-output} tape
%\end{enumerate}

\textbf{Adversary.} The adversary $A\in\ppt$ is an algorithm having four tapes (along with an unknown source of randomness via a random input tape): (1) a read-only {\em adversary-input} tape, (2) a writable {\em query-input} tape, (3) a read-only {\em query-response} tape, and (4) a writable {\em adversary-output} tape

%\begin{enumerate}
%	\item A read-only {\em adversary-input} tape
%	\item A writable {\em query-input} tape
%	\item A read-only {\em query-response} tape
%	\item A writable {\em adversary-output} tape
%\end{enumerate}

\textbf{Simulation.} A simulation is an interactive protocol between the experiment and the adversary when their tapes coincide, and is started by invoking the experiment via the experiment-input tape. % such that the tapes of the experiment and coincide with those of the adversary.

\begin{itemize}
%	
%	The simulation is started by invoking the experiment via the experiment-input tape.
	\item The experiment-input tape contains two inputs: (1) a string of $k$ 1s, where $k$ is a security parameter, and (2) a random string $r$ of $p_{in}(k)$ bits for some $p_{in}\in\poly$. % (specified in $spec$). %This is the input to the simulation.
	\item During the simulation, the experiment and the adversary interact using the common tapes. The adversary terminates after writing a string on the adversary-output tape.
	\item The simulation ends when the experiment writes a \emph{result} on the experiment-output tape. %This is the output of the simulation.
	\item We require the result to be either 0 (indicating $A$ lost) or 1 ($A$ won).
 	\item We denote by $\ms{Expt}_A^{spec}$ the simulation, and by $\ms{Expt}^{spec}_{A}(1^k, r)$ the result when the experiment-input tape contains $(1^k,r)$.%, where $(1^k, r)$ is the  input to the experiment.
%\end{itemize}
%
%\textbf{Simulation.}
	\item Every experiment must be based on the following template:
%\marginpar{Give an example for a CCA2 experiment in the Appendix using our notion}	 	

%\begin{figure}[h]
\begin{enumerate}
	\item $\ms{Expt}_A^{spec}(1^k, r):$
	\item ~~~~~~~~~$\texttt{/* }\mbox{Description of $n$ families }Q_1, Q_2,\ldots, Q_n\in \ptmf\texttt{ */}$
	\item ~~~~~~~~~$\texttt{/* } \mbox{Description of PTM }f:\{0,1\}^{p_{in}(k)}\mapsto \times_{i=1}^{n} \mc{K}_{Q_i}\texttt{ */}$
	\item ~~~~~~~~~$(q_1, q_2, \ldots q_n) \lea f(r)$
	\item \label{inputa}~~~~~~~~~$s\leftarrow A^{Q_1^{q_1}, Q_2^{q_2}, \ldots, Q_n^{q_n}}(1^k,{spec})$
	\item ~~~~~~~~~$\texttt{If (win(}r, QuerySet, s\texttt{)) output 1 else output 0}$
	%\item $\}$
\end{enumerate}

The following discussion is based on the above template.
%We enforce the following requirements:
	\begin{itemize}
	%\item The function $f$ is a deterministic TM that extracts the keys $\{q_i\}_{1\leq i\leq n}$.	
%	\item The parameter $spec$ given to $A$ uniquely identifies the simulation.	\item $A$ interacts with the oracles 	 using the query-input and response tapes.
	\item We do not allow the oracles used by $A$ to maintain state between successive queries\footnote{If state is to be maintained, for instance, each response to the query must use different randomness (and so a query counter must be maintained), then we first assume that adversary can make at most $x$ queries to this oracle, and we replicate the oracle $x$ times, each with different randomness. In the winning condition, we test that each such oracle was queried at most one time.} and assume that a query takes one unit time irrespective of the amount of computation involved.
	\item We require that at any instant $A$ can query at most one oracle. 	
	\item We require that if $r$ is uniformly distributed then so are the keys $q_i~(1\leq i\leq n)$.
	\item The run-time of $A$ is upper-bounded by $p_{run}(k)$ steps for some $p_{run}\in\poly$ (specified in $spec$). 		
	\item $QuerySet$ is a set representing the queries made by $A$ during the simulation. Each element $j$ of this set is an ordered tuple of the type \[(\mathbf{t}_j, \mathbf{i}_j, \mathbf{in}_j, \mathbf{out}_j)\in \mb{N}\times\{1,2,\ldots,n\}\times\{0,1\}^*\times\{0,1\}^*,\]
	indicating respectively, the time, oracle number, input, and the output of each query. %where $i_j\in\{1,2,\ldots, n\}$ indicates the oracle number.	
	\item $\texttt{win}$ is (the PTM description of) an efficiently computable predicate on $(r, QuerySet, s)$.
	\item We say that a family $Q \in {spec}$ if $Q\in \{Q_i\}_{1\leq i\leq n}$.

\end{itemize}
\end{itemize}

\begin{definition} We define \[Adv^{spec}_{A}(k)= \Pr[r\rand\{0,1\}^{p_{in}(k)}:\texttt{Expt}^{spec}_A(1^k, r)=1],\]
the probability taken over the coin tosses of $A$.
\end{definition}

\begin{definition} (\textbf{Obfuscatable family}) For any PTMF $Q_i\in{spec}$, define
\[QuerySet_i=\{(\mathbf{t}_j, \mathbf{i}_j, \mathbf{in}_j, \mathbf{out}_j)|(\mathbf{t}_j, \mathbf{i}_j, \mathbf{in}_j, \mathbf{out}_j)\in QuerySet\wedge \mathbf{i}_j\neq i\}\]
We say that $Q_i$ is \emph{obfuscatable} in $spec$ (written $Q_i\in_{obf} spec$) if
\[\forall r, QuerySet, s:\texttt{\em win}(r, QuerySet, s)=\texttt{\em win}(r, QuerySet_i, s).\]
(In other words, $Q_i\in spec$ is obfuscatable if every element of $QuerySet$ corresponding to oracle $Q_i^{q_i}$ can be removed without affecting the $\texttt{\em win}$ predicate).
%If $Q_i\in {spec}$ is obfuscatable, we write it as $Q_i\in_{obf}spec$.
\end{definition}

\remark{%The notion of white-box security is meaningful only for obfuscatable families.
We claim that it is meaningless to talk about white-box security of specifications where the PTMF to be white-boxed is not-obfuscatable, since it is impossible to keep track of ``queries'' made by an adversary to an obfuscated program. As an example, it is meaningless to talk about obfuscating the decryption oracle of an encryption scheme (or the `signing' oracle of a MAC scheme). %Similarly, it is meaningless to talk about obfuscating the signing oracle in a MAC scheme.
}

An example of a specification for the IND-CCA2 notion of some symmetric encryption scheme is given in Appendix~\ref{ind-cpa-example}.	 	

\subsection{White-box Simulation}
\label{white-box-sim}

Let $\ms{Expt}^{spec}_A$ capture the security of some $spec\in\spec$ (using the template of \S\ref{black-box-sim}). Let $O$ be an obfuscator for some $Q_i$ with $1\leq i\leq n$ such that $Q_i\in_{obf}spec$. Define the corresponding white-box experiment for $(spec, Q_i)$ as follows:
	\begin{enumerate}
	\item $\ms{ExptWB}_{A,O}^{spec, Q_i}(1^k, r):$
	\item ~~~~~~~~~$\texttt{/* }\mbox{Description of $n$ families }Q_1, Q_2,\ldots, Q_n\in \ptmf\texttt{ */}$
	\item ~~~~~~~~~$\texttt{/* } \mbox{Description of PPT }f:\{0,1\}^{p_{in}(k)}\mapsto\times_{i=1}^{n} \mc{K}_{Q_i}\texttt{ */}$
	\item ~~~~~~~~~$(q_1, q_2, \ldots q_n) \lea f(r)$
	\item ~~~~~~~~~$s\leftarrow A^{Q_1^{q_1}, Q_2^{q_2}, \ldots Q_n^{q_n}}(1^k,spec, i, O(Q_i, {q_i}))$
	\item ~~~~~~~~~$\texttt{If (win(}r,QuerySet, s\texttt{)) output 1 else output 0}$
	%\item $\}$
\end{enumerate}

	\begin{itemize}
	\item As before, we bound the running time of $\ms{ExptWB}_{A,O}^{spec,Q_i}$ to $p_{run}(k)$ steps.
	\end{itemize}
\begin{definition} We define \[Advwb^{spec, Q_i}_{A, O}(k)=\Pr[r\rand\{0,1\}^{p_{in}(k)}:\texttt{ExptWB}^{spec, Q_i}_{A, O}(1^k, r)=1],\]
the probability taken over the coin tosses of $A, O$.
\end{definition}

\begin{definition} (\textbf{White-box Property (WBP)}) Let $O$ be an obfuscator for $Q_i\in \ptmf$ and let $spec$ be such that $Q_i\in_{obf}spec$. We say that $O$ satisfies WBP for $(Q_i, {spec})$ if the following holds: 				
\[%\forall A\in\ppt,\exists k'\in \mb{N},\forall k>k': 
\min_{A\in \ppt}\left|Advwb^{spec, Q_i}_{A, O}(k)-Adv^{spec}_{A}(k)\right|\leq negl(|k|).\]

The term $\displaystyle\min_{A\in \ppt}\left|Advwb^{spec, Q_i}_{A, O}(k)-Adv^{spec}_{A}(k)\right|$ is called the \emph{\textbf{white-box advantage} of $A$ w.r.t. $(O,Q_i, spec)$}, and serves a measure of ``useful information leakage'' by the obfuscation.%\footnote{In both the black-box and white-box simulations, the goal of $A$ is to ensure that $\texttt{win}$ evaluates to 1.}

%\[\forall A, \exists k'\in \mb{N},\forall k>k': \left|\begin{array}{c}\Pr[r\rand \{0,1\}^{p_{in}(k)}:\ms{Expt}_A^{spec}(k,r)=1]\\-\Pr[r\rand \{0,1\}^{p_{in}(k)}:\ms{ExptWB}^{spec,i}_{A,O}(k,r)=1]\end{array}\right|\leq negl(|k|),\]
\end{definition}

\begin{definition} (\textbf{Universal White-box Property (UWBP)}) Let $O$ be an obfuscator for $Q\in \ptmf$. We say that $O$ satisfies UWBP for $Q$ if for every $spec\in\spec$ with $Q\in_{obf}{spec}$, $O$ satisfies white-box property for $(Q, spec)$.
\end{definition}
%\begin{lemma}\label{indtouwbp} Let $O$ be an obfuscator that satisfies correctness and soundness w.r.t. IND for some family $Q\in PTMF$. Then $O$ satisfies UWBP for $O$.\end{lemma}

\section{WBC and Obfuscation}
\label{relationships}
%\subsection{Broad relationship between various definitions}
%\marginpar{\amit{
%The following gives an overview of our results.
%\begin{enumerate}
%	\item WBP $\not\Rightarrow$ UWBP. (Theorems~\ref{uwbpapprox} and~\ref{wbpnonlearn})	
%	\item IND $\Rightarrow$ UWBP. (Lemma~\ref{indtouwbp})
%	\item WBP $\not\Rightarrow$ IND. (from 1 and 2 above)%av. case secure obfuscation (Hohenberger {\em et al.})	\item av. case secure obfuscation $\Rightarrow$ white-box property.	\item av. case secure obfuscation $\not\Rightarrow$ universal white-box property.	\item universal white-box property $\stackrel{?}{\Rightarrow}$ av. case secure obfuscation
%%	\item UWBP $\stackrel{?}{\Rightarrow}$ IND.
%%	\item PVBBP $\not\Rightarrow$ WBP.%Relationship with Barak {\em et al.}'s predicate based definition
%%	\item Relationship with UPVBBP and FVBBP.
%\end{enumerate}
%}}

In this section we give some useful relationships between obfuscators, WBP and UWBP. % based on the definitions of \S\ref{obfuscators} and \S\ref{white-box-cryptography}.

\subsection{Negative results}
\label{negative-results}
We note that Barak {\em et al.}'s impossibility results~\cite{Barak2001impossibility} also apply our definitions. In our model, their results can be interpreted as the following:\\

%\begin{proposition}
{\em There exists a pair $(Q, spec)\in \ptmf\times\spec$ with $Q\in_{obf}{spec}$ such that every obfuscator for $Q$ fails to satisfy WBP for $(Q,spec)$.}\\
%\end{proposition}

In other words, there cannot exist a obfuscator that satisfies UWBP for every $Q$. However, their results do not rule out an obfuscator that satisfies the UWBP for some useful family $Q$. We show that even this is not possible unless $Q$ is at least approx. learnable. %This is stated in the next result.

%that there exists a family $Q$ such that for which every obfuscator
%We now give some impossibility results of our definitions.\\

%\textbf{Result 1:} Obfuscators that satisfy universal white-box property for every family do not exist.
%\marginpar{This is Barak's result. Needs to be referred to as such. Perhaps not define it as a result, because it is not a result of this paper.}
%
%\begin{theorem} There exists a family $Q$ and a $spec$ such that $Q\in_{obf}{spec}$ but every obfuscator fails to satisfy white-box property for $(Q, {spec})$. \end{theorem}
%\begin{proof}The proof follows directly from Theorem~\ref{uwbpapprox}.
%\end{proof}

%In other words, there cannot exist a obfuscator that satisfies universal white-box property for every $Q$. What about obfuscators that satisfy universal white-box property for some $Q$? We show next that even this is not possible unless $Q$ is at least approximately learnable.\\

\marginpar{Discussion about \emph{usefulness} in ignore-blob}
\ignore{\color{magenta}Isn't this a bit a weak and ``stupid'' result? I mean, the term spec seems so broad to me at this point that it is impossible to capture this theorem in `real life'; and that it even has no valuable meaning.}
\ignore{\color{red} Well theoretical results have little practical meaning, so practical guys like you will always think like that :-) However, if you look at it clearly, you'll realize that its more meaningful than Barak's original result - They prove that there exists one un-obfuscatable family, while we prove that most useful families cannot be UWBP-obfuscated. True the definition of spec is broad because we want it to capture all possible specs, that humans can imagine in their cryptographic applications. :-)}

\subsubsection{Result 1: No UWBP For ``Interesting'' Families.} (Informal) Obfuscators satisfying UWBP for ``interesting'' families do not exist. More formally,
\begin{theorem}
\label{uwbpapprox}
For every family $Q\in \ptmf\backslash \alf$, there exists a (contrived) $spec\in \spec$ such that $Q\in_{obf}{spec}$ but every obfuscator for $Q$ fails to satisfy the WBP for $(Q, {spec})$.
%
%In other words, the only families that can enjoy UWBP are those that are approximately learnable.
\end{theorem}

\begin{proof} Let $Q\in \ptmf\backslash\alf$. Consider $spec=find$-$q'$ captured below. %by the following simulation.
\begin{enumerate}
	\item $\ms{Expt}_A^{find\mbox{-}q'}(1^k,r:=\la q, q',a\ra):$
	\item $~~~~~~~~~~\texttt{Family }Q(\texttt{Key } q, \texttt{Input } X)\texttt{ \{}$
	\item $~~~~~~~~~~~~~~~~~~\texttt{ /* } \mbox{Description of }Q \mbox{ (used as a black-box)}\texttt{ */}$
	\item $~~~~~~~~~~\}$
	\item $~~~~~~~~~~\texttt{Family }Q_1(\texttt{Key } q_1:=\la q, q',a\ra, \texttt{Input } Y)\texttt{ \{}$
	\item $~~~~~~~~~~~~~~~~~~\texttt{ /* }\mbox{We assume } Y\in \ptm \texttt{ */}$
	%\item $~~~~~~~~~~~~~~~~~~\texttt{ /* We assume } Y\in \ptm \texttt{ */}$
	\item $~~~~~~~~~~~~~~~~~~\texttt{If (}Y(a) = Q^q(a)\texttt{) output }q'\texttt{ else output } 0$
	\item $~~~~~~~~~~~~~~~~~~\texttt{ /* }\mbox{In the above, (the description of) }Q\mbox{ is used as a black-box }\texttt{ */}$
	\item $~~~~~~~~~~~~~~~~~~\texttt{ /* } Y\mbox{ is allowed to run for at most }p(|a|) \mbox{ steps (for some }p\in \poly)\texttt{ */}$
	\item $~~~~~~~~~~\}$
	\item $~~~~~~~~~~\texttt{Function }f(\texttt{Input } r)\texttt{ \{}$
	\item $~~~~~~~~~~~~~~~~~~\texttt{Parse }r \texttt{ as }\la q, q', a\ra$
	\item $~~~~~~~~~~~~~~~~~~\texttt{ /* }\mbox{ we require that }a\in\mc{I}_{Q,|q|} \wedge |q'|=|q|\texttt{ */}$
	\item $~~~~~~~~~~~~~~~~~~\texttt{set } q_1\lea \la q, q', a\ra$
	\item $~~~~~~~~~~~~~~~~~~\texttt{output } q, q_1$
	\item $~~~~~~~~~~\}$
	\item $~~~~~~~~~~q, q_1\lea f(r)$
	\item \label{obforac}~~~~~~~~~~$s\leftarrow A^{Q^{q}_{\phantom{1}},~Q^{q_1}_1}(1^k, find$-$q')$	
	\item ~~~~~~~~~~$\texttt{If (}s=q'\mbox{ and at most one query to } Q_1^{q_1}\texttt{) output 1 else output 0}$
	%\item $\}$
\end{enumerate}
Observe that $Q\in_{obf}{find}$-$q'$. Since $Q\notin \alf$, therefore by virtue of Definitions~\ref{big-defn1}.\ref{alf} and~\ref{correctness-defn}.\ref{apf}, for sufficiently large random $q, q', a$ (and thus, $k$), the following inequalities are guaranteed to hold:
\[\forall A\in \ppt:0\leq Adv^{find\mbox{-}q'}_{A}(k)<\alpha(k)\]
\[\forall A\in \ppt: 1\geq Advwb^{find\mbox{-}q', Q}_{A, O}(k)\geq 1-\beta(k),\]
where $\alpha, \beta$ are negligible functions. %(The constant $1/6$ is arbitrary).
Hence, we have: %\ms{Expt}_A^{find\mbox{-}q'}$ is ``secure'' (see Lemma~\ref{not-alf-implies-cannot-learn}). However, due to Lemmas~\ref{approx-equiv-testable} and~\ref{approx-func-implies-approx-equiv}, every obfuscator will fail to satisfy WBP for $(Q, find$-$q')$ because $O(Q,q)\approx Q^q$ and so $Q_1^{q_1}(O(Q,q))=q'$.
\[\min_{A\in \ppt}\left|Advwb^{find\mbox{-}q', Q}_{A, O}(k)-Adv^{find\mbox{-}q'}_{A}(k)\right|> 1-\alpha(k)-\beta(k),\]
which is non-negligible in $k$. This proves the theorem. %sufficient, since the RHS can be made arbitrarily close to 1/3.
\qed\end{proof}

\ignore{\color{magenta}I'm a little bit skeptical with your proof (and it's my job here to be sceptical ;-)). More in particular with the design of the experiment. To show impossibility of obfuscation of $Q$, you have designed an experiment where the adversary $A$ makes use of an oracle $Q_1$. I find that a bit cheating. It's a bit like proving that encryption is not secure, by enabling the adversary to access the decryption oracle. Try to convince me that it is valid... :-p}
\ignore{\color{red}In a way, I agree, but then this is similar to Barak {\em et al.}'s result, where they prove obfuscator for a certain family doesn't exist by requiring the attacker to output an ``Executable'' version of the target program, which by definition must be satisfied by the obfuscation. IMHO, this is cheating but less cheating than Barak and company :-).}
\remark{\label{prob} The above result applies because the approx. functionality requirement for obfuscators (in the correctness definition of \S\ref{correctness-defn}) is defined only for PTMFs considered as deterministic. What about the extended definitions (such as in~\cite{HofheinzMS07,Hohenberger07re-encryption}) which allow probabilistic families? It turns out that a similar technique can be used for probabilistic families using appropriately extended definitions. This aspect is discussed in \S\ref{pptmfs}.}
\remark{%Result 1 (i.e., Theorem~\ref{uwbpapprox})\label{remark1} is quite strong because it applies to every $Q\notin \alf$. This rules out most interesting families and only leaves various classes of point-functions.
Although we define $\alf$ to be the set of families which can be approximately-learned with a non-negligible advantage (which is quite broad), we note that the above result can be further strengthened by narrowing down the definition of $\alf$ to only families that can be approximately-learned with an overwhelming advantage.} %are to can be extended to a stronger definition of $\alf$ include families that can be also applies In fact, this result can be made stronger

Our next result deals with multiple obfuscations. %First we give a definition.

\subsubsection{Result 2: Simultaneous Obfuscation May Be Insecure.} (Informal) Simultaneous obfuscation of two families may be insecure even if obfuscation of each family alone is secure.
We give a definition before stating this formally.

\begin{definition} (Multiple obfuscations) Let $spec\in \spec$ be the specification defined by $\ms{Expt}^{spec}_A$ using the above template. Let $Q_i, Q_j\in_{obf} spec$ for some $1\leq i, j\leq n$. Let $O$ be an obfuscator for $Q_i, Q_j$.  Extend the white-box simulation $\ms{ExptWB}^{spec,i}_{A,O}$ of \S\ref{white-box-cryptography} by defining a	 corresponding simulation $\ms{ExptWB}^{spec,i,j}_{A,O}$ in which $A$ gets as input in Step 5, the tuple $(1^k, i, j, O(Q_i, q_i),O(Q_j, q_j))$. Finally define,
\[Advwb^{spec, i,j}_{A, O}(k)=\Pr[r\rand\{0,1\}^{p_{in}(k)}:\texttt{ExptWB}^{spec, i,j}_{A, O}(1^k, r)=1],\]
the probability taken over the coin tosses of $A, O$.

We say that $O$ satisfies WBP for $((Q_i, Q_j), spec)$ if the following holds:
\[\min_{A\in \ppt}%forall A\in\ppt,\exists k'\in \mb{N},\forall k>k': 
\left|Advwb^{spec, i,j}_{A, O}(k)-Adv^{spec}_{A}(k)\right|\leq negl(|k|).\]

%We say that $O$ satisfies WBP for $((Q_i, Q_j), spec)$ if the following holds:
%\[\forall A, \exists k'\in \mb{N},\forall k>k': \left|\begin{array}{c}\Pr[r\rand \{0,1\}^{p_{in}(k)}:\ms{Expt}_A^{spec}(k,r)=1]\\-\Pr[r\rand \{0,1\}^{p_{in}(k)}:\ms{ExptWB}^{spec,i,j}_{A,O}(k,r)=1]\end{array}\right|\leq negl(|k|),\]the probability taken over the coin tosses of $O, A$.
\end{definition}

\begin{theorem}
Let $Q_i, Q_j\in \ptmf\backslash\alf$. Then there exists a $spec\in \spec$ with $Q_i,Q_j\in_{obf} spec$ such that even if there exists an obfuscator for $Q_1, Q_2$ satisfying WBP for $(Q_i, {spec})$ and $(Q_j, {spec})$, every obfuscator fails to satisfy WBP for $( (Q_i, Q_j), {spec})$
\end{theorem}
The proof is similar to the proof of Theorem~\ref{uwbpapprox}.

\subsection{Positive Results}
\label{positive-results}

Although the above results rule out the possibility of obfuscators satisfying UWBP for most non-trivial families, they do not imply that a meaningful definition of security for white-box cryptography cannot exist. In fact, any asymmetric encryption scheme can be considered as a white-boxed version of the corresponding symmetric scheme (where the encryption key is also secret). We use this observation as a starting point of our first positive result. A similar observation was used in the positive results of~\cite{HofheinzMS07}.
%\subsubsection{WBP For ``Useful'' Families}
\subsubsection{Result 3: WBP For ``Useful'' Families.} (informal) There exists a non-approx. learnable family (in fact many), and an obfuscator that satisfies WBP for that family under some useful specification. This is stated formally in Theorem~\ref{wbpnonlearn}.
 %Formally,%However there are some differences between their result and ours.

\begin{theorem} \label{wbpnonlearn} Under standard computational assumptions, there exists a pair $(Q,spec)\in \ptmf\backslash \alf\times \spec$ with with $Q\in_{obf} spec$ and an efficient obfuscator $O$ for $Q$ satisfying WBP for $(Q, {spec})$. \end{theorem}

\marginpar{\amit perhaps we don't need to use the specific encryption scheme in the proof of Thm~\ref{wbpnonlearn} if the theorem can be proved based on weaker assumptions such as the existence of one-way functions.}

\marginpar{\amit Also, elaborate why we used this particular scheme instead of, say RSA or Goldwasser Micali encryption scheme. I think there was a very good reason.
}
\begin{proof} We prove this using construction. We will use an encryption scheme based on the BF-IBE scheme~\cite{bonehfranklin}. First we describe a primitive known as a bilinear pairing.  Let $G_1$ and $G_2$ be two cyclic multiplicative groups both of prime order $w$ such that computing discrete logarithms in $G_1$ and $G_2$ is intractable. A bilinear pairing is a map $\hat{e} : G_1 \times G_1 \mapsto G_2$ that satisfies the following properties~\cite{bonehfranklin,boneh03aggregate,boneh01short}.\begin{enumerate}
\item	\emph{Bilinearity}: $\hat{e}(a^x, b^y) = \hat{e}(a, b)^{xy}~\forall a, b \in G_1$ and $x, y \in \mathbb{Z}_w$.
\item \emph{Non-degeneracy}: If $g$ is a generator of $G_1$ then $\hat{e}(g, g)$ is a generator of $G_2$.
\item \emph{Computability}: The map $\hat{e}$ is efficiently computable.
\end{enumerate}
Define a \emph{symmetric} encryption scheme $\mc{E}=(G, E, D)$ as follows.
\begin{enumerate}
	\item \emph{Key Generation ($G$):}  Let $\hat{e}: G_1 \times G_1 \mapsto G_2$ be a bilinear pairing over cyclic multiplicative groups as defined above (such maps are known to exist). Let $|G_1|=|G_2|=w$ (prime) such that $\lfloor\log_2(w)\rfloor=l$. Pick random $g\rand G_1\backslash\{1\}$ and define $\mc{H}:G_2\mapsto\{0,1\}^l$ to be a hash function. Finally pick $x \rand G_1$ and define $key=\la \hat{e}, G_1, G_2, w, g,\mc{H}, x\ra$. The encryption/decryption key is $key$. 	\item \emph{Encryption  ($E$):} The encryption family $E$ using key $key$ is defined as follows. Parse $key$ as $\la \hat{e}, G_1, G_2, w, g,\mc{H}, x\ra$. Let $m\in \{0,1\}^l$ be a message and $\alpha\in \mathbb{Z}_w$ be a random string. Set $(c_1, c_2)\lea E^{key}(x, \alpha)$, where:
	
	\[ E^{key}:\{0,1\}^l\times \mathbb{Z}_w\ni (m, \alpha)\mapsto (\mc{H}(\hat{e}(x^\alpha, g))\oplus m, g^\alpha)\in\{0,1\}^l\times G_1.\] 		 	\item \emph{Decryption ($D$):} The decryption family $D$ using key $key$ is defined as follows. Parse $key$ as $\la \hat{e}, G_1, G_2, w, g,\mc{H}, x\ra$ and compute $m=D^{key}(c_1, c_2)$, where: 	\[D^{key}: \{0,1\}^l\times G_1\ni (c_1, c_2)\mapsto \mc{H}(\hat{e}(c_2, x))\oplus c_1\in \{0,1\}^l.\]
 		
	It can be verified that $D^{key}(E^{key}(m, \alpha))=m$ for valid values of $(m, \alpha)$
\end{enumerate}
The scheme can be proven to be CPA secure if $\mc{H}$ is a random oracle and $w$ is sufficiently large. We construct an obfuscation of the $E^{key}$ oracle that converts $\mc{E}$ into a CPA secure \emph{asymmetric} encryption scheme under a computational assumption.

\textbf{The obfuscator $O$:} The input is $(E, key)$.
\begin{enumerate}
	\item Parse $key$ as $\la \hat{e}, G_1, G_2, w, g,\mc{H}, x\ra$ and set $y\lea\hat{e}(x, g)\in G_2$.
	\item Set ${key}'\lea \la \hat{e}, G_1, G_2, w, g,\mc{H}, y\ra$ and define family $F$ with key ${key}'$ as:
		\[ F^{{key}'}:\{0,1\}^l\times \mathbb{Z}_w\ni(m, \alpha) \mapsto (\mc{H}({y}^\alpha)\oplus m, g^\alpha)\in\{0,1\}^l\times G_1,\]
												where ${key}'$ is parsed as $\la \hat{e}, G_1, G_2, w, g,\mc{H}, y\ra$.

\item Output $F^{{key}'}$.
\end{enumerate}

\begin{claim} $O$ is an efficient obfuscator for $E$ satisfying WBP for $(E, spec)$, where $spec=$ ``IND-CPA security of $\mc{E}$'', assuming that the \emph{bilinear Diffie-Hellman assumption}~\cite{bonehfranklin} holds in $(G_1, G_2)$ and $\mc{H}$ can be considered equivalent to a random oracle.\end{claim}
\begin{proof} We refer the reader to Appendix~\ref{ind-cpa-example} for the formal definition of IND-CPA security. (The IND-CPA game is a restricted version of the IND-CCA2 game given there, by adding the additional check ``no queries to $\mathbf{D}^{key}$'' to the $\texttt{win}$ predicate.)

First note that the obfuscator satisfies correctness for $E$ because $F^{{key}'} = E^{key}$. The proof of the above claim follows from the security of the \textsf{BasicPUB} encryption scheme of~\cite{bonehfranklin}.

\qed\end{proof}

\begin{claim} If $\mc{H}$ is a one-way hash function then $E\in\ptmf\backslash\alf$.\end{claim}
\begin{proof}
Clearly, $F\in \ptmf$ and the following holds:
\[\exists p\in\poly,\forall key\in \mc{K}_{E},\exists {key}'\in \mc{K}_{F}:F^{{key}'}=E^{key}\wedge |{key}'|=|key|+p(|key|).\] By virtue of Lemma~\ref{equivalence-between-non-learn-families}, in order to prove that $E\notin \alf$, it is sufficient to prove that $F\notin \alf$. Finally, it can be proved that if $\mc{H}$ is a one-way hash function then indeed $F\notin \alf$.\footnote{Note that for proving IND-CPA security, we need a stronger assumption on $\mc{H}$, namely that it is equivalent to a random oracle. However, for proving that $E\notin\alf$, the assumption that $\mc{H}$ is a one-way hash function is sufficient.
%We use the following definition of a one-way hash function (based on~\cite{hash-function-defn}).
%
%$H\in \ptmf$ is a {\em one-way function} if $\forall A\in \ptm, \exists k'\in \mb{N},\forall k>k':Adv^{owf}_{A, H}(k)\leq negl(k)$, where, % with key $h$ and let $p_{H}\in poly$
%$Adv^{owf}_{A, H}(k)=\Pr[h,x\rand \{0,1\}^k: H^h(A(1^k,H, h, H^h(x)))=H^h(x)]$.
%
%If $H\in \ptmf$ is a one-way function, then $\forall k,\forall h\rand \{0,1\}^k$, $H^h$ is a {\em one-way hash function}. We define $\mc{H}=H^h$.
}
\qed
\end{proof}
This completes the proof of Theorem~\ref{wbpnonlearn}.
\qed
\end{proof}

\begin{remark}\label{rem3}An interesting observation from Theorem~\ref{wbpnonlearn} is that even though the obfuscator $O$ satisfies WBP for $(E, spec)$, it does not satisfy soundness for $E$ (under Definition~\ref{soundness-defn}). This indicates that the soundness property and WBP are in general independent of each other. %Formally, WBP$\not\Leftrightarrow$ soundness.
\end{remark}

\remark{A reader might wonder why we used the specific encryption scheme in the proof Theorem~\ref{wbpnonlearn}, when we could have used just about any asymmetric scheme (such as RSA), or even the re-encryption scheme of~\cite{Hohenberger07re-encryption}. % or the re-encryption scheme of~\cite{Hohenberger07re-encryption}.
We justify our choice with the following reasons:
\begin{enumerate}
	\item \emph{Why not RSA, El Gamal, etc?}
\begin{enumerate}
	\item Textbook RSA does not enjoy the security notion of IND-CPA. Furthermore, even in RSA variants that are IND-CPA, it is impossible to prove $E\notin\alf$ without relying on additional computational assumptions.
	\item %In the above schemes, proving $E\notin \alf$ is impossible without additional computational assumptions (in fact, 
Encryption in 	El Gamal (and its variants) is learnable.
\end{enumerate}
	\item \emph{Why not re-encryption scheme of~\cite{Hohenberger07re-encryption}?}
\begin{enumerate}
	\item []The obfuscator of~\cite{Hohenberger07re-encryption} does not satisfy approx. functionality as we define. and so their scheme is unsuitable for the proof. (However, the scheme of~\cite{Hohenberger07re-encryption} is the ideal candidate for an analogous example of \S\ref{pptmfs}.)
	%\item The re-encryption scheme of~\cite{Hohenberger07re-encryption} does not reveal the separation between WBP and soundness (Remark~\ref{rem3}). 
%	\item It is possible to modify the above scheme using the Fujisaki-Okamoto transformation~\cite{DBLP:conf/crypto/FujisakiO99} to achieve WBP in the stronger sense of IND-CCA2 security. The same is not known to hold for the re-encryption scheme of~\cite{Hohenberger07re-encryption}. (In the IND-CCA2 game of re-encryption, the adversary is additionally given access to all the decryption oracles with the usual restrictions on decryption queries.)
%%	\item Most importantly, the re-encryption functionality of~\cite{Hohenberger07re-encryption} is learnable if the adversary is in control of the randomness supplied to the obfuscated circuit - a reasonable assumption. (See also Footnote~\ref{hohenberger-randomness} in \S\ref{pptmfs}.) 
\end{enumerate}
\end{enumerate}
}

\subsection{UWBP For Non-Trivial Families}

Let $Q\in \ptmf\cap\lfm$. Then it is easy to construct an obfuscator satisfying UWBP for $Q$ with a non-negligible probability (same as that of  learning $Q$). We call such families \emph{trivial}.

Although Result 1 rules out the possibility of an obfuscator satisfying UWBP for some $Q\in \ptmf\backslash\alf$ (which includes most non-trivial families), it does not rule out the possibility of an obfuscator satisfying UWBP for some non-trivial family $Q\in\ptmf\cap \alf$ (i.e., $Q\in \ptmf\cap\alf\backslash\lfm$). Our next positive result shows that, under reasonable assumptions, this is indeed the case.

\subsubsection{Result 4: UWBP for a non-trivial family.} (informal) There exists an obfuscator satisfying UWBP for a non-trivial (but contrived) family $Q$. Formally, %This is stated formally in Theorem~\ref{main-res2}. %%For simplicity we prove this in the random (permutation) oracle model. However using the results of~\cite{wee}, we can remove the random oracle under reasonable complexity assumptions.

\begin{theorem} \label{main-res2} Under reasonable assumptions, there exists a family $Q\in\ptmf\cap\alf\backslash\lfm$ and an obfuscator $O$ for $Q$ that satisfies UWBP for $Q$.
%\[\forall D,\exists k'\in \mb{N},\forall k\geq k':\left|2\cdot\Pr[q_0,q_1\rand\mc{K}_Q\cap \{0,1\}^{k};b\rand\{0,1\}:D^{Q^{q_b}}(1^k, q_0, q_1)=b]-1\right|\leq negl(k),\]
%Then for $q\rand\mc{K}_Q\cap\{0,1\}^k$, an obfuscator that takes as input $(Q, q)$ and outputs $q$ satisfies IND-soundness for family $Q$. Furthermore the obfuscator also satisfies UWBP for $Q$.
\end{theorem}
\begin{proof} For simplicity, we prove the above result in the random oracle model. Then under the assumption that there exist hash functions equivalent to random oracles, our result can be lifted to the plain model.

Consider the family $Q$ defined below:
\begin{enumerate}
	\item $\texttt{Family }Q(\texttt{Key } q, \texttt{Input } X)\texttt{ \{}$
	\item $~~~~~~~~~~\texttt{ If Random-Oracle}_{|q|}(q||X)=q \texttt{ output }1 \texttt{ else output }0$
	\item $\}$
\end{enumerate}
Here, $\texttt{Random-Oracle}_{|q|}$ is a random oracle mapping arbitrary strings to $|q|$-bit strings. First note that indeed $Q\in \ptmf\cap\alf\backslash\lfm$. It can be proved that $\forall D\in \ppt$ (the distinguisher),\begin{equation}\label{eqn-hash}%\exists k'\in \mb{N},\forall k>k':
\left|\Pr[b\rand\{0,1\}; q_0,q_1\rand \{0,1\}^k\cap\mc{K}_Q:D^{Q^{q_b}}(1^k, q_0,q_1)=b]- \frac{1}{2}\right|\leq negl(k),\end{equation} the probability taken over the coin tosses of $D$.
For any $k$, let $q\rand \{0,1\}^k\cap\mc{K}_Q$. Consider an obfuscator $O$ that takes in as input $(Q, q)$ and simply outputs a description of $Q^q$ as the obfuscation of $Q^q$. Let $spec\in\spec$ be such that $Q\in_{obf}spec$ but $O$ does not satisfy WBP for $(Q, spec)$ w.r.t. some adversary $A\in \ppt$. If $A$ has a non-negligible white-box advantage w.r.t. $(O, Q, spec)$, then $A$ can be directly converted into a distinguisher $D$ such that Equation~\ref{eqn-hash} does not hold, thereby arriving at a contradiction.
%has a non-negligible white-box advantage w.r.t. $(O, Q, spec)$, then $A$ is essentially a distinguisher  obfuscation.
%First note that $O$ satisfies IND-soundness for $Q$ is easy to see. For contradiction assume there exists a $spec \in \spec$ with $Q\in_{obf}spec$ such that $O$ does not satisfy WBP for $(Q, spec)$. Let $A$ be the adversary thats fails the WBP. Using $A$ we construct a distinguisher $D$ such the inequality in the theorem does not hold. %It is easy to see that the obfuscator satisfies IND-soundness. %Use Wee's construction of obfuscation under random permutation assumption. Use the above obfuscation as the starting family and then construct an obfuscator for this obfuscated family. (``double obfuscation'') then show that if there exists $spec \in \spec$ s.t. UWBP fails for $(Q,spec)$ and $Q\in_{obf}\spec$, then $A$ can be used by $D$ to distinguish the two keys $q_0,q_1$ in the theorem. Basically, this can be an extension of the ``distinguishing attack property''. Every such spec is simulatable by $D$. Hence every $\spec$ will satisfy this property. The result follows from Lemma~\ref{dap}. as a black-box, we can construct another algorithm $D\in\ppt$
\qed\end{proof}

\section{The Case Of Probabilistic PTMFs}
\label{pptmfs}
In this section, we consider probabilistic (i.e., randomized) functions based on the definitions of~\cite{HofheinzMS07,Hohenberger07re-encryption}. In contrast to conventional constructions of PTMFs (such as the encryption algorithm of the probabilistic encryption scheme in the proof of Theorem~\ref{wbpnonlearn} and Appendix~\ref{ind-cpa-example}), where randomness is considered as part of the input tape, the definitions of~\cite{HofheinzMS07,Hohenberger07re-encryption} consider randomness as part of the key tape.\footnote{\label{hohenberger-randomness}In the construction of~\cite{Hohenberger07re-encryption}, the PTMF has additional randomness on the input tape (a.k.a. `re-randomization values', which are supplied by the adversary). We ignore this additional randomness in our discussion (adversary cannot be trusted to supply randomness), since we are focusing on the security of the \emph{obfuscator of some given PTMF}, and not of the \emph{specification in which the PTMF is used}.} We call a PTMF of the latter type, a probabilistic PTMFs (PPTMF).

%\footnote{Defining randomness as part of the key tape does not seem to give any practical advantage, except perhaps the fact that Proposition~\ref{if-ind-then-alf} does not hold anymore.}

Intuitively, a PPTMF is simply an ordinary PTMF $Q$ with part of the key used for randomness, so that two different keys are ``equivalent'' provided only their random bits are different.

Formally, a PPTMF is any pair of the type $(Q,\tau)$, where $Q\in \ptmf$ and $\tau$ is an equivalence relation on $\mc{K}_Q$ that partitions $\mc{K}_Q$ into equivalence classes, s.t. \[\forall q_1, q_2\in \mc{K}_Q: \tau(q_1,q_2)=1 \iff \mbox{only the random bits of $q_1, q_2$ are different.}\]
We denote the set of all PPTMFs by $\pptmf$.
\begin{definition}
\label{big-defn}
In the following, let $(Q,\tau)\in \pptmf$. % and $\tau$ be an equivalence relation on $\mc{K}_Q$.

\begin{enumerate}
	\item  Let $q\in \mc{K}_Q$. Then:
\begin{itemize}
	\item For any $(a, z)\in \mc{I}_{Q,|q|}\times\{0,1\}^*$, we say that $z$ is \emph{$\tau$-equal} to $Q^{q}(a)$ (written $z=_\tau Q^{q}(a)$) if
					\[\exists q'\in \mc{K}_Q:z=Q^{q'}(a)\wedge \tau(q,q')=1.\]
	\item  For any $X\in \tm$ we say that $X$ is \emph{$\tau$-equal} to $Q^q$ (written $X=_\tau Q^q$) if \[\forall a\in\mc{I}_{Q,|q|}: X(a)=_\tau Q^{q}(a).\]
%	\item  For any $X\in \tm$ we say that $X$ is \emph{$\tau$-approx. equal} to $Q^q$ (written $X\approx_\tau Q^q$) if
%					\[\forall \ell:\Pr[a\rand \{0,1\}^\ell:X(a)=_\tau Q^{q}(a)]\geq \frac{2}{3}.\]
\end{itemize}

%\[\Pr[a\rand \{0,1\}^l:X(a)\notin\{Q^{q'}(a)| q'\in \mc{K}_Q\wedge \tau(q)=\tau(q')\}]\leq negl(l).\]

	\item  We define a \emph{$\tau$-(approx.) learnable} family by replacing ``$=$'' with ``$=_\tau$'' in the definition of (approx.) learnable families.
	The following claim is easy to prove. % (the converse is not true).
	
	\begin{claim} If $Q$ is not $\tau$-approx. learnable then $Q\notin\alf$.
	\end{claim}
	
	%Note that unlike $=$ and $\approx$, $=_\tau$ and $\approx_\tau$ are not equivalence relations for TMs.% for some $q'$ s.t. $\tau(q)=\tau(q')$.''

	\item \label{pptmf-obf}Let $O:\ptmf\times\{0,1\}^*\mapsto \tm$ be a randomized algorithm. Then:
		
\begin{itemize}
	\item $O$ satisfies \emph{$\tau$-approx. functionality} for $Q$ if
					\[\forall q\in \mc{K}_Q,\forall a \in\mc{I}_{Q,|q|}:\Pr[O(Q, q)(a)\neq_\tau Q^{q}(a)]\leq negl(|q|),\]
		%\[\forall  q\in \mc{K}_Q,\forall a\in \{0,1\}^*:\Pr[O(Q, q)(a)\notin\{Q^{q'}(a)| q'\in \mc{K}_Q\wedge \tau(q)=\tau(q')\}]\leq negl(|q|+|a|),\]
		the probability taken over the coin tosses of $O$.

	\item $O$ satisfies \emph{$\tau$-correctness} for $Q$ if Definition~\ref{correctness-defn} (of \S\ref{correctness}) holds when ``approx. functionality'' is replaced by ``$\tau$-approx. functionality''.
	
	\item	$O$ is a \emph{$\tau$-obfuscator} for $Q$ if it satisfies $\tau$-correctness for $Q$.% and soundness for $Q$ (\S\ref{soundness}, Definition~\ref{soundness-defn}).
\end{itemize}
	\item\label{tau-decidable} $Q$ is \emph{$\tau$-decidable} if there exist $p,p'\in\poly$ and for every $k\in\mb{N}$, there exists an efficiently computable map \[\mathbf{f}_k:\{0,1\}^{p(k)}\mapsto  \{0,1\}^{k}\cap\mc{K}_Q\times \ptm,\] such that for all $(q,Z)\lea\mathbf{f}_k(r)$, the following holds:
		\begin{itemize}
			\item $\forall a,z\in\{0,1\}^*: Z(a,z)=1\iff z=_\tau Q^q(a)$.
			\item If $r$ is uniformly distributed then so is $q$.
			\item $|Z|\leq p'(k)$.
		\end{itemize}
\end{enumerate}
\end{definition}
%Lemma
%%~\ref{pptmfs-approx-equiv-testable} %and
%~\ref{analog-approx-func-implies-approx-equiv} below %are
%is the equivalent of Lemma %~\ref{approx-equiv-testable} and~
%\ref{approx-func-implies-approx-equiv} %respectively
%for PPTMFs. The proof is left as an exercise.% Definition~\ref{big-defn}.
%
%
%%\begin{lemma}\label{pptmfs-approx-equiv-testable} Let $X\in\ptm$ and $(Q,\tau)\in\pptmf$ such that $Q$ is $\tau$-decidable under Definition~\ref{big-defn}, Item~\ref{tau-decidable}. Then for every $k\in \mathbb{N}$, and every $(q,Z)\leftarrow \textbf{\em f}_k(r)$ with $r\in\{0,1\}^{p(k)}$, there exists a polynomial (in $k$) algorithm that uses $X$ and $Z$ as black-boxes and decides with overwhelming (in $k$) probability if $X\stackrel{?}{\approx}_\tau Q^q$. %The algorithm need not know the description of $Q$.
%%\end{lemma}
%
%
%\begin{lemma}
%\label{analog-approx-func-implies-approx-equiv}
%If $O$ is a $\tau$-obfuscator for $Q$ then $\forall q:O(Q, q)\approx_\tau Q^q$. %The converse is not true.
%\end{lemma}%
%%However, we use a different (and slightly restricted) definition, which is more useful in practice. %this definition leaves a lot to be desired. For instance, how to define random bits? We use a slightly different definition.
%

We claim that in any meaningful PPTMF construction $(Q,\tau)$, the family $Q$ must be $\tau$-decidable (this is true for the constructions of ~\cite{HofheinzMS07,Hohenberger07re-encryption}).
Theorem~\ref{uwbpapprox1} below is the equivalent of Theorem~\ref{uwbpapprox} for PPTMFs. The proof follows directly %from Lemma %~\ref{pptmfs-approx-equiv-testable} and~
%\ref{analog-approx-func-implies-approx-equiv}
after replacing ``$=$'' with ``$=_\tau$'' in the proof of Theorem~\ref{uwbpapprox}.

\begin{theorem}
\label{uwbpapprox1}
For every $(Q,\tau)\in \pptmf$ with $Q$ $\tau$-decidable but not $\tau$-approx. learnable, there exists $spec\in \spec$ such that $Q\in_{obf}{spec}$ but every $\tau$-obfuscator for $Q$ fails to satisfy the WBP for $(Q, {spec})$.
%
%In other words, the only families that can enjoy UWBP are those that are approximately learnable.
\end{theorem}
\subsection{An Open Question: WBP and Soundness}
%<Section removed. Need to incorporate ideas about PPPTMs here.. i.e., try to relate WBP and IND-soundness for an obfuscator satisfying $\tau$-approx. functionality. See \S\ref{pptmfs}. >

\label{wbpfromobf}
Let $((Q, \tau),spec)\in \pptmf\times\spec$ such that the following is true:

\begin{enumerate}
	\item $Q$ is not $\tau$-approx. learnable.
	\item $Q\in_{obf}spec$
	\item $Q$ is $\tau$-decidable
	\item $O$ is a $\tau$-obfuscator for $Q$ satisfying IND-soundness (\S\ref{soundness}, Definition~\ref{soundness-defn}).
\end{enumerate}
A useful question is: \emph{Given $spec$, can we decide if $O$ satisfies WBP for $(Q,spec)$?}
%\footnote{Interestingly, we do not have to worry about whether $spec$ is `meaningful' w.r.t. the relation $\tau$ or not, since both WBP and soundness are defined independently of $\tau$ (although it makes sense to worry only about $spec$s that have some relevance w.r.t. $\tau$).}

\emph{Why is it useful?} Ideally, we would like the WBP to be satisfied. However, WBP is defined w.r.t. a family and a specification while soundness is defined w.r.t. a family, independent of the specification. On the one hand, due to this simplified definition, obfuscator designers may find it appealing. On the other hand, it is possible that IND-soundness may be too strong to be satisfied even though WBP is (w.r.t. to some specification), as in the example in the proof of Theorem~\ref{wbpnonlearn}. Nevertheless, we consider it an interesting question to characterize cases when the WBP can be reduced to IND-soundness. The assumption that $O$ is a $\tau$-obfuscator (rather than an obfuscator) for $Q$ is necessary to falsify Proposition~\ref{ind-to-alf}, which rules out interesting families.

\emph{An Open Question:} If $spec$ is such that the only oracle available to $A$ is the one being obfuscated, then the WBP for $spec$ indeed holds if IND-soundness holds. The result also holds if $spec$ has additional oracles that always output the same string (which can be given as an auxiliary input to the distinguisher - cf. ``distinguishable attack property'' of~\cite{Hohenberger07re-encryption}). At this stage, an open question is:  \emph{how to characterize $spec$s where $A$ is given additional oracles which output query-dependent strings?}

\section{Conclusion}

%{\amit We initiated a formal study of WBC by defining WBP and UWBP. We gave some negative results relating obfuscators and UWBP. We also a positive result for WBP. As some open questions: we have the following:

%(a) Thm 4 is true ? |b) Can similar methods be developed for PVBBP as for IND in the above section.}
%

In this work, we initiated a formal study of White-Box Cryptography (WBC) and investigated its relationship with obfuscation. We presented definitions and (im)possibility results for obfuscators of {\em specific} classes of `interesting' programs families - those that are not approx. learnable. The security requirements of WBC is captured by means of a \emph{White-Box Property (WBP)}, which is defined for some scheme and a security notion (we call the pair a specification). The security requirement of an obfuscator is captured using a \emph{soundness} property. We showed that WBP and soundness are in general quite independent of each other by giving some examples where one is satisfied but the other is not.

Although the WBP is defined for a particular (family, specification) pair, soundness is only defined for a given family and is independent of the specification. A natural question is whether there exist non-trivial families for which the WBP w.r.t. every specification can be reduced to the soundness of an obfuscator for that family. Loosely speaking, an obfuscator that achieves this is said to satisfy the \emph{Universal White-Box Property (UWBP)} for that family. We showed that the UWBP fails for every family that is not approx. learnable. However, we show that under reasonable assumptions there exists an obfuscator $O$ satisfying UWBP for a non-learnable but approx. learnable family. %At this stage it is an open question whether such families exist. %, the UWBP is too much to ask for. %we showed that this is impossible. %where soundness alone is sufficient to prove WBP for every specification
Furthermore, the specification we used for our negative result is quite contrived. Hence, it seems reasonable to expect that a meaningful notion of security for WBC based on WBP can still be achieved for ``normal'' specifications. As a possible example of this, we presented a (non-trivial) non-approx. learnable family for which there does exist an obfuscator satisfying the WBP in a real-world specification. Additionally, we showed that there exists a (contrived) family $Q\in \alf\backslash\lfm$ for which there exists an obfuscator satisfying UWBP for $Q$.

$~$\\
\textbf{Acknowledgement:} 
%The work described in this document has been partly financially supported by the IAP Programme P6/26 BCRYPT of the Belgian State (Belgian Science Policy), and in part by the European Commission through the IST Programme under Contract IST-021186-2 for the RE-TRUST project.

We are thankful to Amir Herzberg for useful discussions and for motivating the topic. We would also like to thank Dennis Hofheinz, Gregory Neven and Ioannis Atsonios for their valuable feedback.
\bibliographystyle{plain}
\bibliography{refs}
\appendix
%\newpage
\textbf{APPENDIX}%
\section{An Example Specification}
\label{ind-cpa-example}

Let $\mc{E}=(G, E, D)$ be a symmetric encryption scheme. We define the IND-CCA2 specification using the following simulation. The corresponding specification is called $ind\mbox{-}cca2\mbox{-}\mc{E}$. In the following, the key generation algorithm, $G$ takes in as input the security parameter $(1^k)$ and a $k$ bit random string $\gamma$. It outputs a $k$ bit encryption/decryption key $key$.%, which is the same for both encryption and decryption (as in a symmetric encryption scheme).

\begin{enumerate}
	\item $\ms{Expt}_{A,O}^{ind\mbox{-}cca2\mbox{-}\mc{E}}(1^k, r):$
	\item ~~~~~~~~~$\texttt{Family }\mathbf{E}(\texttt{key }key,\texttt{Input } \la \alpha,m\ra) \texttt{ \{}$
	\item ~~~~~~~~~~~~~~~~~~~~~~~~$\texttt{/* } \alpha\mbox{ is randomness}\texttt{ */}$
	\item ~~~~~~~~~~~~~~~~~~$\texttt{output }E(key, \alpha, m)$
	\item ~~~~~~~~~$\}$
	\item ~~~~~~~~~$\texttt{Family }\mathbf{D}(\texttt{key }key,\texttt{Input } c) \texttt{ \{}$
	\item ~~~~~~~~~~~~~~~~~~$\texttt{output }D(key, c)$
	\item ~~~~~~~~~$\}$
	\item ~~~~~~~~~$\texttt{Family }\mathbf{C}(\texttt{key }\la b,key, \beta\ra ,\texttt{Input } \la m_0,m_1\ra) \texttt{ \{}$
	\item ~~~~~~~~~~~~~~~~~~~~~~~~$\texttt{/* }\mathbf{C} \mbox{ is the challenge oracle. } b \in \{0,1\}\mbox{ is a bit. } \beta \mbox{ is randomness}\texttt{ */}$
	\item ~~~~~~~~~~~~~~~~~~$\texttt{output }E(key, \beta, m_b)$
	\item ~~~~~~~~~$\}$
	\item ~~~~~~~~~$\texttt{Function }f(\texttt{Input } r) \texttt{ \{}$
	\item ~~~~~~~~~~~~~~~~~~~~~~~~$\texttt{/* } f:\{0,1\}^{2k+1}\mapsto\{0,1\}^k\times\{0,1\}^k\times\{0,1\}^{2k+1}\texttt{ */}$
	\item ~~~~~~~~~~~~~~~~~~$\texttt{parse }r\texttt{ as }\la \gamma,\beta, b\ra$
	\item ~~~~~~~~~~~~~~~~~~$key\lea G(1^{|\gamma|}, \gamma)$
	\item ~~~~~~~~~~~~~~~~~~$\texttt{output }key, key, \la b, key, \beta\ra $
	\item ~~~~~~~~~$\}$	
	\item ~~~~~~~~~$key, key, \la b, key, \beta\ra \lea f(r)$
	%exttt{/* } \mbox{Description of PPT }f:\{0,1\}^{p_{in}(k)}\mapsto\times_{i=1}^{n} \mc{K}_{Q_i}\texttt{ */}$
	\item ~~~~~~~~~$s\leftarrow A^{\mathbf{E}^{key}, \mathbf{D}^{key}, \mathbf{C}^{\la b, key, \beta\ra}}(1^k,ind\mbox{-}cca2\mbox{-}\mc{E})$
	\item ~~~~~~~~~$\texttt{If (win(}r,QuerySet, s\texttt{)) output 1 else output 0}$
	%\item $\}
\end{enumerate}
Here $\texttt{win}:=$ ``If (At most one query to $\mathbf{C}^{\la b,key, \beta\ra})~\wedge$ (No query to $\mathbf{D}^{key}$ on output of $\mathbf{C}^{\la b,key, \beta\ra}$ after query to $\mathbf{C}^{\la b,key, \beta\ra})\wedge (s=b)$''.

Clearly, $\mathbf{E}\in_{obf}ind\mbox{-}cca2\mbox{-}\mc{E}$.
\end{document}